\documentclass[]{aastex631}

\usepackage{gensymb}

\newcommand{\revision}{}

\accepted{\today}

\submitjournal{PSJ}

\shorttitle{The SAMOSA Intercomaprison Protocol}
\shortauthors{Haqq-Misra et al.}

\begin{document}

\title{The Sparse Atmospheric MOdel Sampling Analysis (SAMOSA) intercomparison: Motivations and protocol version 1.0. \\
A CUISINES model intercomparison project.}

\correspondingauthor{Jacob Haqq-Misra}
\email{jacob@bmsis.org}

\author[0000-0003-4346-2611]{Jacob Haqq-Misra}
\affiliation{Blue Marble Space Institute of Science, Seattle, WA, USA}

\author[0000-0002-7188-1648]{Eric T. Wolf}
\affiliation{Laboratory for Atmospheric and Space Physics, University of Colorado Boulder, Boulder, CO, USA}

\author[0000-0002-5967-9631]{Thomas J. Fauchez}
\affiliation{NASA Goddard Space Flight Center, Greenbelt, MD 20771, USA}
\affiliation{American University, Washington DC, USA}

\author[0000-0002-7086-9516]{Aomawa L. Shields}
\affiliation{Department of Physics and Astronomy, University of California, Irvine, CA, USA}

\author[0000-0002-5893-2471]{Ravi K. Kopparapu}
\affiliation{NASA Goddard Space Flight Center, Greenbelt, MD 20771, USA}

\begin{abstract}
Planets in synchronous rotation around low-mass stars are the most salient targets for current ground- and space-based missions to observe and characterize. Such model calculations can help to prioritize targets for observation with current and future missions; however, intrinsic differences in the complexity and physical parameterizations of various models can lead to different predictions of a planet's climate state. Understanding such model differences is necessary if such models are to guide target selection and aid in the analysis of observations. This paper presents a protocol to intercompare models of a hypothetical planet with a 15 day synchronous rotation period around a 3000\,K blackbody star across a parameter space of surface pressure and incident instellation. We conduct a sparse sample of 16 cases from a previously published exploration of this parameter space with the ExoPlaSim model. By selecting particular cases across this broad parameter space, the SAMOSA intercomparison will identify areas where simpler models are sufficient as well as areas where more complex GCMs are required. Our preliminary comparison using ExoCAM \revision{shows} general consistency between the climate state predicted by ExoCAM and ExoPlaSim except in regions of the parameter space most likely to be in a \revision{steam atmosphere or incipient} runaway greenhouse state. We use this preliminary analysis to define \revision{several} options for participation in the intercomparison by models of all levels of complexity. The participation of other GCMs is crucial to understand how the atmospheric \revision{states across this parameter space differ} with model capabilities. 
\end{abstract}

\section{Introduction} \label{sec:intro}

Planets orbiting M-dwarf stars are ideal targets for detection with missions such as TESS \citep{ricker2014transiting} and PLATO \citep{rauer2014plato}, while JWST \citep{gardner2006james} and the next generation of space telescopes \citep{tumlinson2019next} will provide opportunities for the characterization of terrestrial planet atmospheres in such systems. Terrestrial planets around low-mass stars that could retain surface liquid water on their surfaces are considered to be ``habitable'' planets and represent ideal targets to search for spectroscopic signs of life. M-dwarf systems also comprise about 75\% of the total population of main sequence stars, so the characterization of such systems will also remain a high priority for exoplanet science in general \citep[e.g.,][]{tarter07,seager2013exoplanet,shields2016habitability,meadows2018factors}.

Many planets in such systems are expected to be in synchronous rotation around their host stars, so that one side of the planet experiences perpetual daylight and the opposing hemisphere resides in perpetual night. Numerous climate models have been developed for understanding this unique climate configuration and its impact on the habitability of planets orbiting M-dwarf stars \citep[e.g.][]{joshi1997simulations,joshi2003climate,merlis2010atmospheric,carone2014connecting,kopparapu:2017,turbet2016habitability,del2019habitable}. The ability of such models to provide meaningful constraints that can guide observations depends in part on identifying areas where multiple models show robust agreement and areas where predictions differ. \revision{One example of such a model intercomparison was conducted by \citet{yang2019simulations}, which found that model differences in the treatment of radiative transfer and cloud parameterization can lead to significant differences for planets in synchronous rotation around M-dwarf stars, compared to planets in rapid rotation around G-dwarf stars. Another} example of such an effort was the TRAPPIST-1 Habitable Atmosphere Intercomparison (THAI, \cite{fauchez2020trappist}), which compared four general circulation models (GCMs) and several other models \revision{(including energy balance models and radiative-convective equilibrium models)} using predetermined benchmark climate configurations for TRAPPIST-1e \citep{fauchez2021trappist,turbet2021trappist1,sergeev2021trappist1}. The \revision{efforts by both \citet{yang2019simulations} and} THAI illustrated the need for continued exoplanet model intercomparisons, which can identify areas for improving \revision{individual models} and also begin to approach model ensemble predictions to guide exoplanet observations. 

A more recent effort at systematically developing a framework for exoplanet model intercomparisons began with the CUISINES (Climates Using Interactive Suites of Intercomparisons Nested for Exoplanet Studies) workshop in 2021 \citep{cuisinesreport}. CUISINES was organized to support modeling intercomparisons for the exoplanet community, with THAI as the first example. The goal of CUISINES is to provide a framework for conducting exoplanet model intercomparisons and to provide benchmarks to the community to help testing and validating exoplanet models. In this paper, we present one such intercomparison protocol for models of habitable planets in M-dwarf systems.

The purpose of the Sparse Atmospheric MOdel Sampling Analysis (SAMOSA) intercomparison is to compare GCMs of varying complexity, as well as other climate models, across a broad parameter space of atmospheric nitrogen pressure and stellar instellation. By selecting particular cases across this broad parameter space, the SAMOSA intercomparison will identify areas where simpler models are sufficient as well as areas where more complex GCMs are required. The identification of such areas of the parameter spaces may depend on the GCM used and on how far from the standard modern Earth atmosphere the simulation takes place. The design of this intercomparison is motivated by a recent study that used the ExoPlaSim GCM of intermediate complexity to conduct hundreds of simulation of an Earth-sized planet in a fixed synchronous orbit around an M-dwarf host star \citep{paradise2021exoplasim}. ExoPlaSim is a computationally efficient GCM that can rapidly explore a large parameter space that would be difficult or intractable with a GCM of full complexity. Comparing ExoPlaSim results can be instead accomplished by using sparse sampling methods to select a small number of cases from the complete set to be simulated with a full GCM.

This paper defines the protocol for the SAMOSA intercomparison. We describe the sampling method and the parameters for the sparse sample of 16 cases in section \ref{sec:sample}. However, we also recognize that at least some of these cases may be in \revision{an extremely} hot regime that may not be easily simulated by many models. \revision{In some instances, this difficultly is the result of a limited scope of the model physics, such as the use of a radiative transfer scheme that is only validated up to a fixed temperature threshold. In other instances, warm atmospheres with significant water vapor may enter a dynamical regime that leads to inaccuracies in the representation of atmospheric dynamics. Such problems may be addressable through additional model development, which, if implemented, may result in stable ``steam atmosphere'' solutions \citep[e.g.][]{turbet2021day}. Yet in other instances, a planet may enter a true runaway greenhouse state \citep[e.g.][]{ingersoll1969runaway,goldblatt:2013,leconte2013increased} in which the accumulation of water vapor increases the greenhouse effect until the planet's oceans evaporate. Numerical simulation of runaway greenhouse atmospheres is untenable for most climate models; at best, such models can demonstrate a trend toward \revision{an incipient} runaway greenhouse state before the model becomes numerically unstable. In order to demonstrate the difficultly of simulating these various hot atmospheres, we} therefore also present a set of simulations of these 16 cases using the ExoCAM GCM \citep{Wolf_2022} in section \ref{sec:exocam}. This preliminary comparison shows regions of the ExoPlaSim parameter space that are likely to be \revision{numerically} unstable \revision{hot or} runaway atmospheres, while other regions of the parameter space show better agreement between ExoPlaSim and ExoCAM. We refrain from providing any further quantitative comparison between ExoCAM and ExoPlaSim, as such analysis will be saved for a future, multi-model intercomparison paper.  

We invite GCMs, energy balance models (EBMs), radiative-\revision{convective} equilibrium (RCE) models, and others that may be relevant to participate in the SAMOSA intercomparison. We offer \revision{several} different options for participation that range from a single model run to the full set of 16 cases, \revision{as well as additional sequences that extend coverage of the parameter space, which are all} described in section \ref{sec:options}. The SAMOSA intercomparison will continue for the next 2-3 years, with the goal of providing a comprehensive assessment of the range and unity of exoplanet climate models.

\section{Defining the Sparse Sample}\label{sec:sample}

We begin by describing our approach for defining the sparse sample across a parameter space of pressure and instellation. This sparse sample will form the basis of the SAMOSA intercomparison.

\citet{paradise2021exoplasim} used the ExoPlaSim general circulation model (GCM) of intermediate complexity to study the effect of varying background pressure and stellar instellation on climate. Their set of model calculations assumed an Earth-sized planet with aquaplanet surface conditions and a fixed 15-day synchronous rotation around a 3000\,K blackbody host star. The model atmospheres included a fixed 400\,ppm of CO$_2$ with no ozone or other trace gases \revision{but with H$_2$O produced from surface evaporation}. The parameter space consisted of a grid of 460 simulations with ExoPlaSim that spans 0.1--10\,bars for surface pressure (represented \revision{mainly} as the N$_2$ gas pressure) and 400--2600\,W\,m$^{-2}$ for instellation. Further details about the model configuration were described by \citet{paradise2021climate}, who performed a similar set of experiments with ExoPlaSim but with a Sun-like host star.

Intermediate GCMs like ExoPlaSim are computationally efficient and capable of exploring a much broader parameter space than more complex GCMs. \revision{For example, ExoPlaSim cases take on the order of minutes to complete, whereas a model like ExoCAM can require 35,000 core hours per simulation, which takes about 2 weeks on a supercomputing cluster. Even with the capability of running multiple simulations in parallel, the use of a full GCM like ExoCAM requries a significant investment in researcher time to initialize the set of simulations, inspect and restart crashed cases, conduct mid-run analyses, and manage the model output data (up to $\sim$80\,GB per simulation). The} computational advantages of ExoPlaSim are offset by simplifications in the model physics, most notably with the model radiative transfer that includes two shortwave bands and one longwave band. Full GCMs, by contrast, may have dozens or more bands each for shortwave and longwave regions. \revision{For example, ExoCAM uses 37 bins for the longwave calculation from wavenumber infinity to 4,030\,cm$^{-1}$ and 53 bins for the shortwave calculation from 875\,cm$^{-1}$ to 42,087\,cm$^{-1}$.} ExoPlaSim also considers water vapor as a minor atmospheric constituent, which may limit the accuracy of its predictions for warm climates \revision{where water vapor becomes a major constituent}. These and other simplifications in ExoPlaSim will lead to differences in the steady-state climate predictions when compared to full GCMs, but the magnitude of these differences may vary across the parameter space.

Directly comparing the complete set of 460 ExoPlaSim cases to a full GCM would be computationally \revision{expensive}, but such comparisons can be instead accomplished by using sparse sampling methods to select a small number of cases from the complete set to be simulated with a full GCM. \citet{paradise2020large} emphasized the need for such comparisions by calling for ``studies which sparsely re-sample the PlaSim-surveyed parameter space with higher-complexity GCMs,'' which would ``be useful in verifying PlaSim’s results, as well as in helping to indicate where PlaSim is inaccurate or likely has missing physics (such as a dynamic ocean or sea ice drift).'' This model protocol paper represents an attempt to conduct such a sparse sampling comparison between ExoPlaSim and other models.

We use a quasi-Monte Carlo approach to select a sparse sample of cases from the set of 460 ExoPlaSim simulations. Quasi-Monte Carlo methods enable improved uniformity when sampling from a large distribution by making use of low-discrepancy sequences (i.e., randomly selecting cases or samples that are uniformly distributed over a large parameter space). Conventional Monte Carlo methods, by contrast, rely on pseudorandom number generators that may have poor space-filling properties (i.e., higher discrepancy) for a small number of samples. For further discussion about quasi-Monte Carlo methods, see \citet{lemieuxmonte} and references therein. Our particular approach uses a scrambled Sobol sequence to select cases \citep{joe2003remark,joe2008constructing}, which requires the number of points to be equal to a power of 2 (i.e., the number of points selected in each uniformly distributed sequence must be 2$^n$, where $n$ can be any \revision{whole number}).

\begin{figure}[ht!]
\centering
\includegraphics[width=1.00\linewidth]{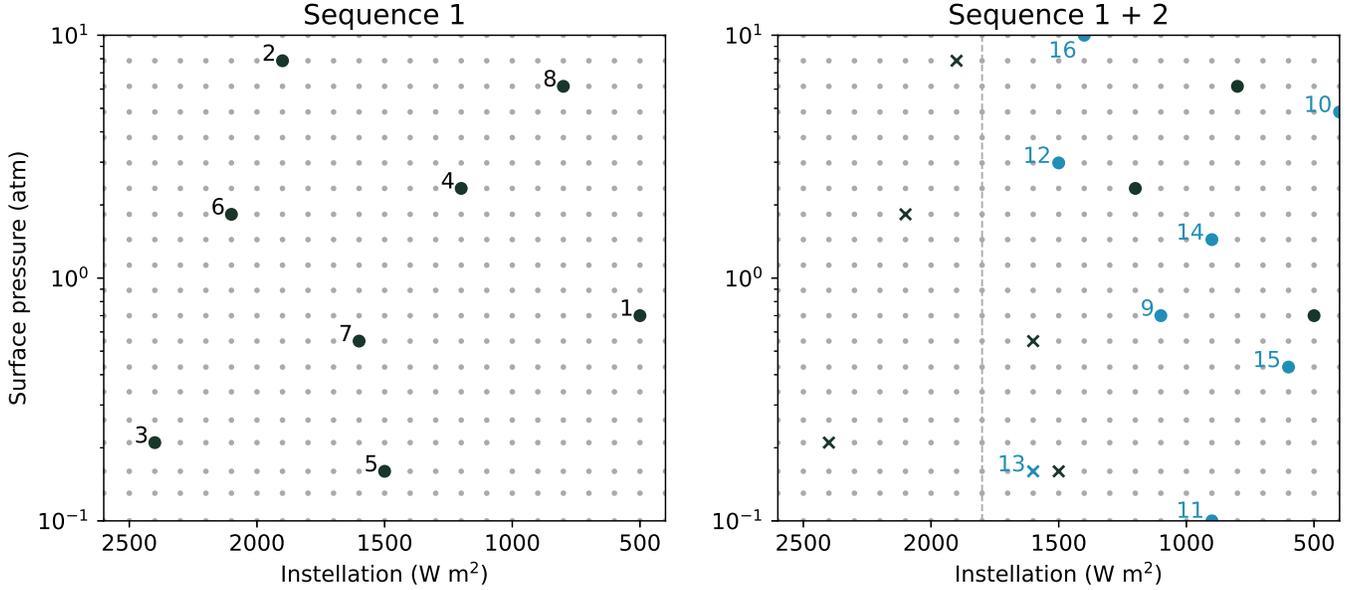}
\caption{Quasi-Monte Carlo selection of 16 points from the ExoPlaSim parameter space, with \textit{Sequence 1} (left) consisting of 8 points that span the full parameter space and \textit{Sequence 2} (right) consisting of 8 points within a region where full GCMs are expected to be numerically stable. Points marked with X indicate \revision{numerically unstable} atmospheres when calculated with ExoCAM. Numbered labels correspond to the sample number in Table \ref{tab:variables}.\label{fig:sobol}}
\end{figure}

We show the selection of a sequence of 8 cases from the ExoPlaSim parameter space in the left panel of Figure \ref{fig:sobol} and top half of Table \ref{tab:variables}, which we refer to as \textit{Sequence 1}. The distribution of \textit{Sequence 1} can be visually inspected to show reasonable uniform properties, with the value of the centered discrepancy for this sequence equal to 0.00781. The seed value for this sequence is fixed at 5936744 for reproducibility. The purpose of \textit{Sequence 1} is to provide a sparse selection of cases to compare with GCMs, or other models, that span the full parameter space. Comparing these 8 cases with ExoPlaSim and more complex GCMs will identify the broad regions of this parameter space that show the greatest extent of agreement or disagreement.

\begin{table}[ht!]
\centering
\caption{Variables from the Quasi-Monte Carlo selection of cases in \textit{Sequence 1} and \textit{Sequence 2}, which are also shown in Figure \ref{fig:sobol}.\label{tab:variables}}
\begin{tabular}{cccc}
    & Sample & Instellation (W\,m$^{-2}$) & Surface pressure (bar)  \\
\hline\hline
\textit{Sequence 1}    & 1      &  500      & 0.70                   \\
                       & 2      &  1900     & 7.85                   \\
                       & 3      &  2400     & 0.21                   \\                       
                       & 4      &  1200     & 2.34                   \\
                       & 5      &  1500     & 0.16                   \\
                       & 6      &  2100     & 1.83                   \\
                       & 7      &  1600     & 0.55                   \\                       
                       & 8      &  800      & 6.16                   \\
\hline                       
\textit{Sequence 2}    & 9      &  1100     & 0.70                   \\
                       & 10     &  400      & 4.83                   \\
                       & 11     &  900      & 0.10                   \\                       
                       & 12     &  1500     & 2.98                   \\
                       & 13     &  1600     & 0.16                   \\
                       & 14     &  900      & 1.44                   \\
                       & 15     &  600      & 0.43                   \\
                       & 16     &  1400     & 10.0                   \\
\hline                       
\end{tabular}
\end{table}

We also select a second sequence of 8 cases from a narrower set of the ExoPlaSim parameter space. This approach is known as importance sampling \citep{andral2022attempt} and provides a way to use quasi-Monte Carlo methods to sample from a subspace that is expected to be physically interesting. In this case, the ExoPlaSim results shown in Figure 25 by \citet{paradise2021exoplasim} indicate numerical instabilities or runaway/steam atmospheres at high pressure and instellation, which may be a result of some of the physical parameterizations in ExoPlaSim. Such warm climates can also be challenging for full GCMs to accurately represent. We therefore construct \textit{Sequence 2} by sampling only from 400--1800\,W\,m$^{-2}$ for instellation, shown in the bottom half of Table \ref{tab:variables}. The distribution of \textit{Sequence 2} is shown in the right panel of Figure \ref{fig:sobol} as light blue dots, with the darker dots indicating the cases from \textit{Sequence 1}. The centered discrepancy of \textit{Sequence 2} is 0.00817, and the seed value is 8398110.

The cases selected by \textit{Sequence 1} and \textit{Sequence 2} represent only 3.5\% of the total number of simulations in the ExoPlaSim parameter space, but the quasi-Monte Carlo approach provides confidence that these sequences give representative behavior across the full parameter space and within the subsampled region. The use of only 16 cases across both sequences is tractable even for computationally intensive models and can facilitate a comparision between ExoPlaSim and more complex full GCMs. We will use these 16 cases as the basis for the SAMOSA intercomparison. \revision{In Section \ref{sec:options}, we also provide additional options for extending beyond this set of 16 cases to give better coverage of the parameter space.}

\section{Preliminary Comparison}\label{sec:exocam}

We next perform calculations with ExoCAM \citep{Wolf_2022} to simulate the cases in \textit{Sequence 1} and \textit{Sequence 2} as shown in Table \ref{tab:variables}, using the planetary parameters listed in Table \ref{tab:parameters}. These calculations are intended to provide an initial assessment of the regions of parameter space where ExoPlaSim and ExoCAM differ significantly. We do not perform any detailed quantitative analysis and focus on identifying the regions of parameter space where complex GCMs like ExoCAM are likely to be numerically stable and computationally feasible.

\begin{table}[ht!]
\centering
\caption{Fixed stellar and planetary parameters. Earth values are assumed to be the mean for each parameter.\label{tab:parameters}}
\begin{tabular}{cc}
\hline
Star and Spectrum & 3000\,K blackbody \\
Rotation period & 15\,d \\
Orbital period & 15\,d \\
Planet mass & 1\,M$_\Earth$ \\
Planet radius & 1\,R$_\Earth$ \\
Planet density & $\rho_\Earth$ \\
Gravity & 1\,g$_\Earth$ \\
Atmospheric CO$_2$ & 400\,ppm \\
Surface conditions & aquaplanet \\
\hline                       
\end{tabular}
\end{table}

ExoCAM is configured with 4$\degree\times$5$\degree$ horizontal resolution with 40 vertical layers extending 3 orders in magnitude in dry atmospheric pressure.  Because each simulation has a different assumed dry surface pressure, all simulations also have a different model top.  We assume a thermodynamic ocean with a 50-meter depth and no \revision{zonal or meridional} ocean heat transport\revision{; this ``slab ocean'' configuration has been used in other exoplanet modeling studies with ExoCAM \citep[e.g.,][]{kumar2016inner,adams2019aquaplanet,komacek2019atmospheric,rushby2020effect}}.  Sea ice and snow albedos have been weighted against the input spectra (3000~K blackbody) to yield an accurate representation in two-channel form \citep[e.g.,][]{shields:2018}.  Surface albedos \revised{are divided into two} channels, with sea ice albedos calculated to be 0.65 and 0.17 and snow albedos calculate to be 0.97 and 0.46, in visible and near-infrared channels, respectively.  Water vapor adds mass to the atmosphere and virtual temperature is used to account for atmospheric density changes due to moistening atmospheres.  Here, we use the finite volume dynamical core \citep{lin&rood:1996} along with the CAM4 set of moist physics routines \citep{rasch&kristjansson:1998}. We use the ExoRT radiation model \textit{n68equiv}, which features absorption by H$_2$O, CO$_2$ using HITRAN2016 \citep{gordon:2017} and MT$\_$CKD version~3.2 for the H$_2$O continuum \citep{clough:2005}.  Other collision induced absorptions are included, but are not relevant in this study.  More details on ExoCAM can be found in \revision{the paper by} \citet{Wolf_2022} and related references to the National Center for Atmospheric Research Community Earth System Model contained within.  

\begin{table}[ht!]
\centering
\caption{ExoCAM and ExoPlaSim global mean climate statistics and general climate categorization for \textit{Sequence 1} and \textit{Sequence 2}.\label{tab:exocam}}
\begin{tabular}{cccccccc}
Sample & Model & Climate State & T$_{S,mean}$(K) & T$_{S,max}$(K) & T$_{S,min}$(K) & Ice Fraction & Stratospheric Water (kg\,kg$^{-1}$)  \\
\hline\hline
1    & ExoCAM     &  \textit{snowball} & 196.8  & 268.9  & 147.4  & 1.0  & 3.7$\times$10$^{-8}$  \\
     & ExoPlaSim  &  \textit{snowball} & 176.0  & 264.7  & 124.6  & 1.0  & 1.8$\times$10$^{-15}$  \\
\hline                      
2    & ExoCAM     &  \textit{unstable}   & -      & -      & -      & 0.0 & - \\
     & ExoPlaSim  &  \textit{hot}       & 368.2  & 377.3  & 318.2  & 0.0  & 3.0$\times$10$^{-2}$  \\
\hline
3    & ExoCAM     &  \textit{unstable}   & -      & -      & -      & 0.0 & - \\
     & ExoPlaSim  &  \textit{hot}       & 296.6  & 320.5  & 259.5  & 0.0  & 5.3$\times$10$^{-1}$  \\
\hline
4    & ExoCAM     &  \textit{warm}       & 260.0  & 299.6  & 211.1  & 0.64 & 1.2$\times$10$^{-6}$ \\
     & ExoPlaSim  &  \textit{warm}       & 254.0  & 300.4  & 200.6  & 0.7 & 4.2$\times$10$^{-10}$ \\
\hline
5    & ExoCAM     &  \textit{unstable}   & -      & -      & -      & 0.0 & - \\
     & ExoPlaSim  &  \textit{warm}      & 265.7  & 306.1  & 219.0  & 0.6  & 6.9$\times$10$^{-3}$ \\
\hline
6    & ExoCAM     &  \textit{unstable}   & -      & -      & -      & 0.0  & - \\
     & ExoPlaSim  &  \textit{hot}       & 343.1  & 352.2  & 333.9  & 0.0   & 1.0$\times$10$^{-1}$  \\
\hline
7    & ExoCAM     &  \textit{unstable}   & -      & -      & -      & 0.0  & - \\
     & ExoPlaSim  &  \textit{warm}      & 279.7  & 316.8  & 245.5  & 0.5   & 1.5$\times$10$^{-4}$ \\
\hline
8    & ExoCAM     &  \textit{icy}       & 243.8  & 286.4  & 218.2  & 0.82  & 2.4$\times$10$^{-9}$  \\
     & ExoPlaSim  &  \textit{icy}       & 215.9  & 274.5  & 158.1  & 0.95  & 8.4$\times$10$^{-15}$ \\
\hline                       
\hline                      
9    & ExoCAM     &  \textit{icy}      & 244.8  & 286.4  & 200.4  & 0.72 & 1.2$\times$10$^{-5}$ \\
     & ExoPlaSim  &  \textit{icy}      & 239.9  & 301.7  & 182.1  & 0.7  & 3.5$\times$10$^{-9}$ \\
\hline  
10   & ExoCAM     &  \textit{snowball} & 194.1  & 230.0 & 166.5 & 1.0  & 1.6$\times$10$^{-12}$  \\
     & ExoPlaSim  &  \textit{snowball} & 172.8  & 232.7 & 132.2 & 1.0  & 4.4$\times$10$^{-17}$  \\
\hline 
11   & ExoCAM     &  \textit{icy}      & 234.0  & 291.0 & 195.2 & 0.78 & 2.8$\times$10$^{-3}$   \\  
     & ExoPlaSim  &  \textit{icy}      & 211.3  & 289.7 & 133.5 & 0.9  & 2.8$\times$10$^{-4}$  \\
\hline  
12   & ExoCAM     &  \textit{hot}      & 350.9  & 355.7 & 348.2 & 0.0  & 4.4$\times$10$^{-4}$   \\
     & ExoPlaSim  &  \textit{hot}      & 345.7  & 355.0 & 337.2 & 0.0  & 2.9$\times$10$^{-4}$  \\
\hline  
13   & ExoCAM     &  \textit{unstable}  & -      & -      & -      & 0.0 & -  \\
     & ExoPlaSim  &  \textit{warm}     & 272.9  & 309.5  & 229.0  & 0.6  & 1.0$\times$10$^{-1}$ \\
\hline  
14   & ExoCAM     &  \textit{icy}      & 236.8  & 288.7 & 183.2 & 0.74  & 2.4$\times$10$^{-7}$  \\
     & ExoPlaSim  &  \textit{icy}      & 224.5  & 297.5 & 158.1 & 0.8   & 3.8$\times$10$^{-12}$ \\
\hline  
15   & ExoCAM     &  \textit{icy}      & 211.5  & 285.6 & 157.9 & 0.88  & 4.7$\times$10$^{-7}$  \\
     & ExoPlaSim  &  \textit{icy}      & 186.3  & 287.0 & 122.5 & 0.95  & 3.3$\times$10$^{-13}$ \\
\hline  
16   & ExoCAM     &  \textit{hot}      & 356.7  & 360.7 & 353.8 & 0.0   & 1.5$\times$10$^{-7}$  \\
     & ExoPlaSim  &  \textit{hot}      & 346.3  & 356.0 & 306.3 & 0.0   & 5.8$\times$10$^{-7}$ \\
\hline                       
\end{tabular}
\end{table}

In Table \ref{tab:exocam} we summarize the climate results from ExoCAM.  Cases 2, 3, 5, 6, 7, and 13 enter numerically unstable states (marked with X on fig. \ref{fig:sobol}). In each of these cases we find large positive top-of-atmosphere energy imbalances combined with steep increases in the global mean temperature \revision{and stratospheric specific humidity}.  Model limitations, both numeric and functional, prevent us from exploring to terminal states \revision{for such hot atmospheres that are} characterized by steam atmospheres and $\sim$1500~K surface temperatures \citep{goldblatt:2013}. Cases 3, 5 and 13 were terminated due to numerical instabilities between radiation and dynamics coupling, an unfortunately common problem for low-pressure, high-instellation, non-dilute water-vapor atmospheres in our model.  Cases 2, 6, and 7 are numerically stable, but increasing temperatures push the model out-of-bounds of the radiative transfer absorption coefficient tables (\revision{which fixes such cases to} maximum values of 500~K) and also run afoul of the wet-bulb temperature parameterization used in the cloud model.  Both of these aspects are in theory fixable, but here we stop the model integration when these limits are reached.   

\begin{figure}[ht!]
\centering
\includegraphics[width=1.0\linewidth]{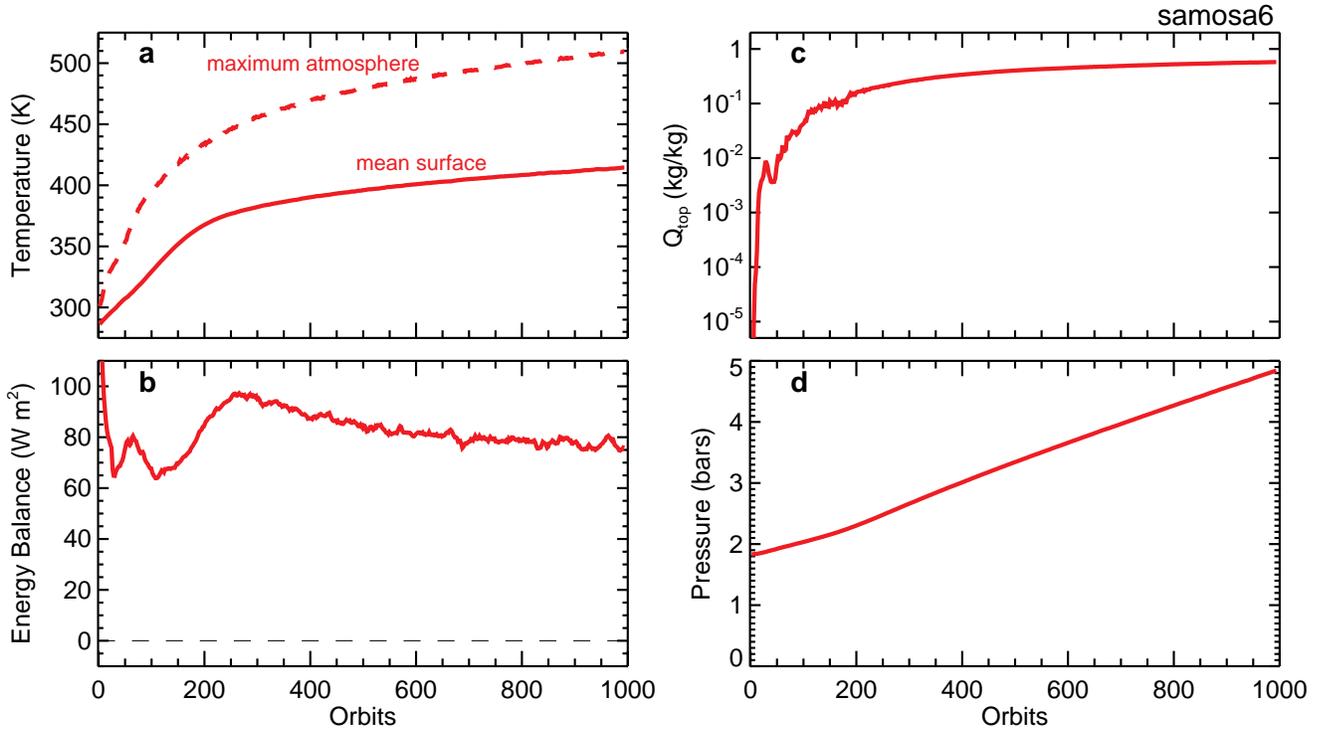}
\caption{\revision{ExoCAM time} series simulation output from case 6, showing mean (solid) and maximum (dashed) temperature (top left), stratospheric specific humidity (top right), net energy balance (bottom left), and total pressure (bottom right).\label{fig:samosa_runaway}}
\end{figure}

Figure \ref{fig:samosa_runaway} shows time series model outputs from case 6, \revision{which is an example of a} case \revision{in the onset of an incipient runaway greenhouse}.  At the time of model termination, the mean surface temperature was $\sim$415~K, with maximum atmosphere temperatures found at $\sim$510~K.  The top-of-atmosphere energy imbalance is $\sim$80$\%$ Wm$^{-2}$, and the trend of gradual continued warming looks to persist for an undetermined amount of model time.  Water vapor has become a major constituent of the atmosphere, with the total atmospheric surface pressure reaching $\sim$5~bars, while the dry pressure at the start of the simulation was only 1.83~bars.  Even at the upper most grid-box of the model, specific humidity \revision{exceeds the 10$^{-3}$\,kg\,kg$^{-1}$ threshold that corresponds to rapid stratospheric water vapor loss}.  Analysis and demonstration of the transient behavior of our simulations as shown in Figure \ref{fig:samosa_runaway}, along with our previous model experience by \citet{kopparapu:2017} give us confidence in declaring these cases to be incipient runaway greenhouse worlds.

\begin{figure}[ht!]
\centering
\includegraphics[width=1.0\linewidth]{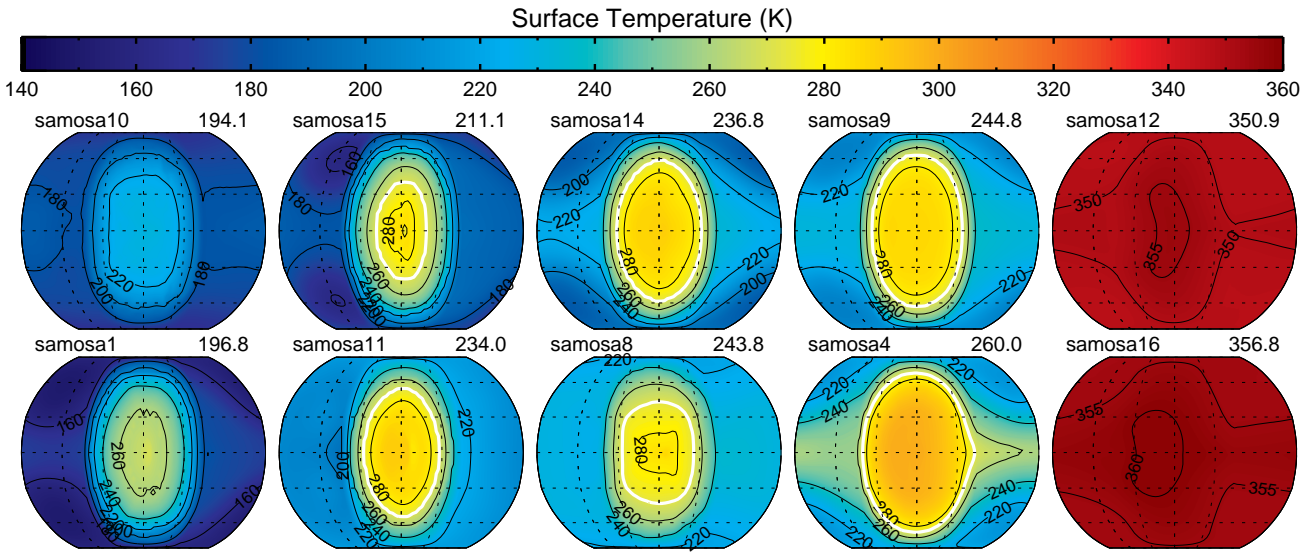}
\caption{Average surface temperature for the 10 cases from \textit{Sequence 1} and \textit{Sequence 2} that remain stable with ExoCAM.\label{fig:ts_contour}}
\end{figure}

The remaining cases yield stable climates, ranging from snowball planets to hot, moist worlds.  Figure \ref{fig:ts_contour} shows the surface temperature contour plots from the 10 simulations that yield climatologically stable states. Note that we have sorted the figures in ascending order of mean surface temperature, with the coldest case in the upper left corner and warmest non-runaway case in the lower right corner. We have also categorized the climate state of each case in Table \ref{tab:exocam} according to its global mean temperature as \textit{hot} (300-400\,K), \textit{warm} (250-300\,K), \textit{icy} (200-250\,K), or \textit{snowball} (below 200\,K). The corresponding climate state and global mean climate statistics for ExoPlaSim is also shown in Table \ref{tab:exocam}. We note that cold ice-dominated simulations (70+$\%$ ice coverage) have a small residual negative surface energy imbalance and cold temperature drift even after 200 model years of simulation.  A closer examination reveals that while day-side temperatures and sea-ice fractions have long since stabilized, the night-side cold traps continue to gradually cool as the sea ice sheet continues to thicken.  For computational expediency we have truncated these runs after 150 model years.  Cases 12 and 16 were found to stabilize in a hot and most climate state, with surface temperatures above 350\,K.  

\revision{The stratospheric water vapor content is also shown in Table \ref{tab:exocam} for both models, which is calculated as the specific humidity at the model top. For ExoPlaSim, the model top is fixed at 50\,hPa for all simulations. For ExoCAM, the model top is variable and equal to 3 orders of magnitude lower than the surface pressure. In all cases, the ExoCAM stratospheric water vapor content is greater than that calculated by ExoPlaSim, with the hot cases 12 and 16 showing the closest agreement between the two models. Although the difference in the model top may account for some of the differences between these cases, these differences primarily arise from the fact that ExoCAM treats water vapor as a major atmospheric constituent whereas ExoPlaSim does not. The results in Table \ref{tab:exocam} only include the stable ExoCAM solutions, but incipient runaway cases also show a rapid increase in stratospheric water vapor; for example, panel c of Figure \ref{fig:samosa_runaway} shows the model top water vapor evolution in the unstable ExoCAM case 6, where the stratospheric specific humidity is 5.8$\times$10$^{-1}$\,kg\,kg$^{-1}$ and climbing.}

\begin{figure}[ht!]
\centering
\includegraphics[width=1.0\linewidth]{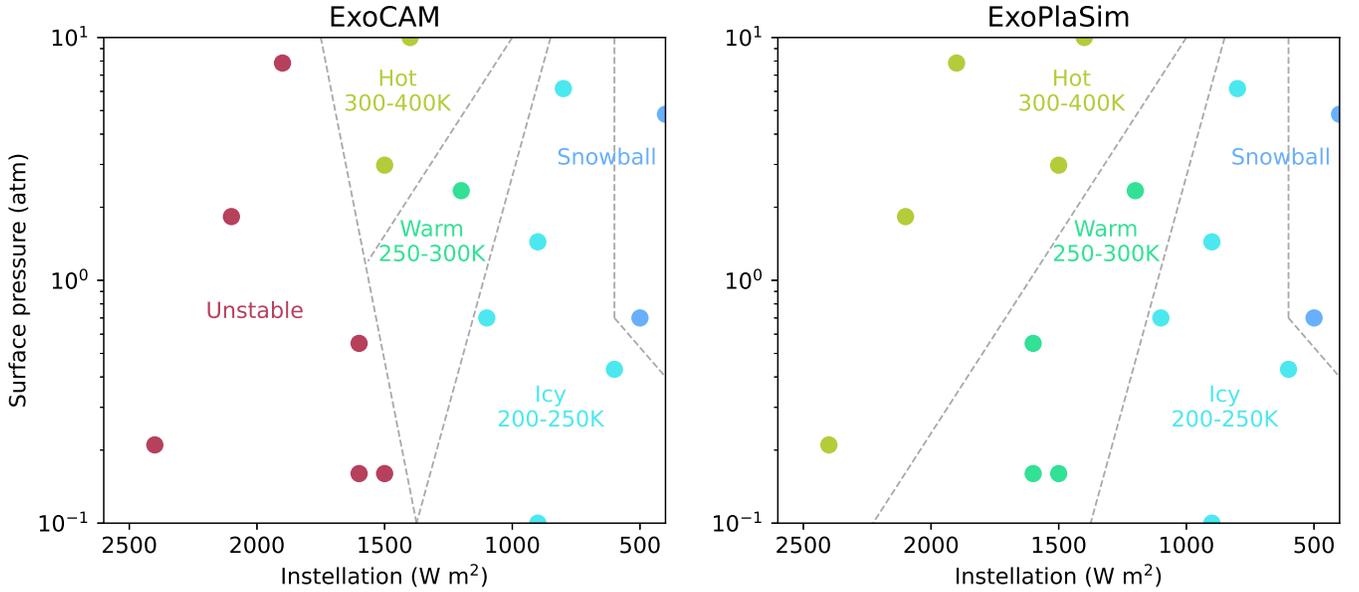}
\caption{Global average surface temperature for all 16 cases using ExoCAM (left) and ExoPlaSim (right). Colors correspond to the climate state calculated for each case, with dashed lines showing approximate boundaries between climate states.\label{fig:ts_grid}}
\end{figure}

Figure \ref{fig:ts_grid} plots all of our 16 cases from \textit{Sequence 1} and \textit{Sequence 2} with simulations from ExoCAM in the left panel and ExoPlaSim in the right panel. Each point is color-coded to indicate a particular climate state, with dashed lines showing approximate boundaries between climate states. Simulations in \revision{a numerically unstable} regime only occur for ExoCAM, with ExoPlaSim calculating these to be in hot (cases 2, 3, 6) or warm (cases 5, 7, and 13) climate states. Such behavior is expected, as \citet{paradise2021exoplasim} noted that they ``have not validated ExoPlaSim in moist greenhouse and runaway greenhouse regimes'' and chose to ``include these warmer models for the sake of completeness.'' \revision{Although ExoPlaSim shows numerically stable solutions for these extremely hot atmospheres, it is worth noting that \citet{paradise2021exoplasim} calculate the stratospheric water vapor content of these cases to be near or in excess of 10$^{-3}$\,kg\,kg$^{-1}$, which is indicative of a moist or runaway greenhouse state.} Case 4 is the only warm climate state calculated with ExoCAM, and the region of warm parameter space predicted by ExoCAM is smaller than for ExoPlaSim. The icy and snowball climate states of ExoCAM and ExoPlaSim otherwise remain consistent with each other. 

This preliminary comparison between ExoCAM and ExoPlaSim highlights the similarities between the climate states predicted between the two models across most of the parameter space, aside from the region prone to a runaway greenhouse or numerical instability; however, further intercomparison with other models, and GCMs in particular, is needed to understand the extent to which intermediate complexity models like ExoPlaSim give similar results to more complex GCMs. We do not conduct any further detailed analysis in this protocol paper, but we note that the results shown in Table \ref{tab:exocam} indicate notable differences between ExoCAM and ExoPlaSim when individual cases are examined. The contribution of additional simulations by other GCMs is essential for understanding whether these differences result from model complexity, the choice of physical parameterizations, or other model-dependent factors. 

\revision{We also note that subsequent analysis should also compare other factors that will be useful in diagnosing the regions of agreement or disagreement between different models. Intercomparison with models of varying complexity, such as radiative-convective equilibrium models, may reveal areas of parameter space where simpler models can make robust predictions about the onset of moist or runaway greenhouse states based on the stratospheric water vapor content. Dynamical circulation patterns may also be a useful diagnostic for comparing various GCMs across the full parameter space, as differences in physical parameterizations and the dynamical core may result in different predictions for the atmosphere's dynamical regime \citep[e.g.][]{haqq2018demarcating}. These and other atmospheric diagnostics will be of critical importance for understanding the role of different GCMs and models of varying complexity in making useful predictions about climate across this parameter space.}

Although the results shown in Figure \ref{fig:ts_grid} (left panel) may seem to indicate a narrow region of parameter space for warm planets, it is important to emphasize that these simulations all use a fixed amount of CO$_2$. This implicitly ignores the carbon-silicate weathering cycle, which could act to regulate climate and produce warmer states with higher CO$_2$ at lower instellation values.  We have also omitted ocean heat transport, which for continent-free planets could lead to considerable warming and de-icing of the night-side on colder tidally-locked worlds \citep[e.g.,][]{hu2014role,yang2014water,yang2019ocean,checlair2019no,kane:2021}. Furthermore, we note that in this preliminary analysis we have \revision{made} an approximate delineation between five climate states, which reveals that useful information can be deduced using a spare grid representation.  However, one of the challenges of using sparse grid techniques is to devise scientifically driven interpolation methods to stitch the the sparse grids together, which is non-trivial given the inherent complexity and non-linearity of tipping points in planetary climate. We will explore such analysis methods in our future intercomparison paper.

\section{Participation Options}\label{sec:options}

We invite other climate models to join the SAMOSA intercomparison. Participation is open to GCMs, EBMs, RCE models, and any others that can examine terrestrial climate for a synchronously rotating planet as a function of surface pressure and instellation. Participating models should be configured with the fixed parameters listed in Table \ref{tab:parameters}. 

\begin{figure}[ht!]
\centering
\includegraphics[width=1.00\linewidth]{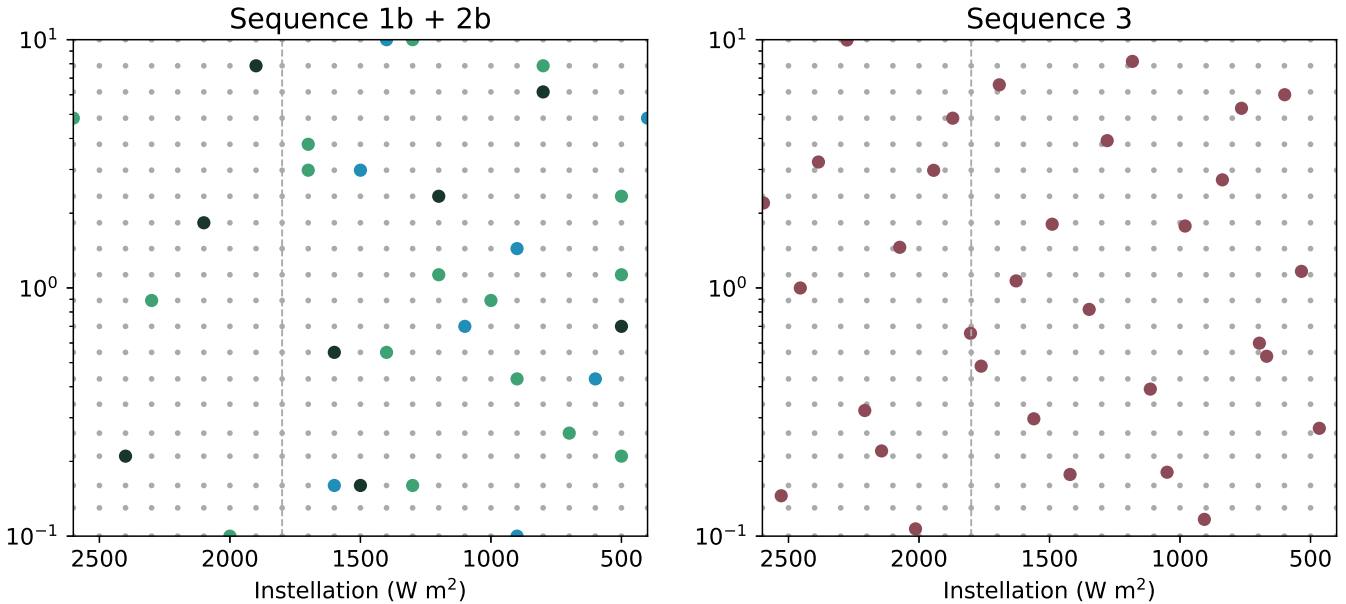}
\caption{Additional quasi-Monte Carlo selection of points from the ExoPlaSim parameter space. The left panel shows 16 additional points (green dots) that extend the selection shows in Figure \ref{fig:sobol}, with \textit{Sequence 1b} consisting of 8 additional points continuing from \textit{Sequence 1} (dark dots), which spans the full parameter space, and \textit{Sequence 2b} consisting of 8 additional points continuing from \textit{Sequence 2} (blue dots), which is restricted to a region where full GCMs are expected to be numerically stable. The right panel shows \textit{Sequence 3}, which is a sample of 32 new points that are selected within the parameter space but do not necessarily fall along the ExoPlaSim grid.\label{fig:extended}}
\end{figure}

\revision{The cases in \textit{Sequence 1} and \textit{Sequence 2} are listed in Table \ref{tab:variables} and plotted in Figure \ref{fig:sobol}, which provide coverage across the full parameter space; however, the use of additional cases could provide even greater insight into the similarities and differences in the climate states predicted by different models. We therefore provide additional sequences that allows participants to choose to extend the number of simulations for comparison with ExoPlaSim and other participating models. The first option extends \textit{Sequence 1} by 8 additional points to give \textit{Sequence 1b}, which provides additional cases across the full parameter space (and reduces the discrepancy to 0.00225). The second option extends \textit{Sequence 2} by 8 additional points to give \textit{Sequence 2b}, which provides additional cases across the region of parameter space that is expected to be numerically stable for GCMs (and reduces the discrepancy to 0.00212). The variables for \textit{Sequence 1b} and \textit{Sequence 2b} are given in Table \ref{tab:variablesext} and plotted in the left panel of Figure \ref{fig:extended} as green dots. A third option provides a new sparse sample across the full parameter space, \textit{Sequence 3}, which consists of 32 points that do not necessarily fall along the predefined ExoPlaSim grid points. The centered discrepancy of \textit{Sequence 3} is 0.000604, and the seed value is 1043337. The variables for \textit{Sequence 3} are given in Table \ref{tab:variablesext} and plotted in the right panel of Figure \ref{fig:extended}. These optional cases are intended to supplement the 16 primary cases that were discussed in Section \ref{sec:exocam}.}

We offer several options for participation in SAMOSA, each with a different number of required simulations:
\begin{enumerate}
    \item Simulate only the \textit{warm} ExoCAM case (1 model run, case 4)
    \item Simulate only the cases in \textit{Sequence 2} (8 model runs, cases 9-16)
    \item Simulate all stable ExoCAM cases (10 model runs, cases 1,4,8-12,14-16)
    \item \revision{Simulate all cases in \textit{Sequence 1} and \textit{Sequence 2} (16 model runs, cases 1-16)}
    \item \revision{Simulate all cases in \textit{Sequence 1}, \textit{Sequence 1b}, \textit{Sequence 2} and \textit{Sequence 2b} (32 model runs, cases 1-32)}
    \item \revision{Simulate all cases in \textit{Sequence 1}, \textit{Sequence 1b}, \textit{Sequence 2}, \textit{Sequence 2b}, and \textit{Sequence 3} (64 model runs, cases 1-64)}
\end{enumerate}
These options are intended to provide flexibility for participating models, given the constraints imposed by the computational requirements for a model integration, the physical parameterizations within the model, and the availability of researcher time. The parameter space studied here can push many models into extreme limits, as shown by our ExoCAM results, so we do not prescribe a specific timestep, resolution, or any other model requirements for this intercomparison. Any parameters not listed in Tables \ref{tab:variables} and \ref{tab:parameters} should be selected at the discretion of the researcher based on prior experience with the particular model.

\revision{We note that the activities involved in scientific research, including computational modeling, can serve as sources of carbon emissions. \citet{aujoux2021assess} have provided a framework for assessing the largest emissions factors for a particular scientific project, which include professional travel, digital communication, computational simulations, data transmission and storage, and the use and transportation of hardware. Most, if not all, of these factors can contribute to the total emissions of a project like SAMOSA. The choice of a researcher to perform additional computational simulations, preserve excess data in cloud storage, or engage in travel for conferences or collaboration is a complex cost-benefit analysis that may include subjective elements on the part of the researcher. Quantifying these sources of carbon emissions may be useful as researchers contemplate such questions. In this protocol paper, we do not perform any such quantification of emissions associated with simulations or other factors, but such factors may be relevant for individuals and research groups as they decide which of the SAMOSA participation options to choose.}

\begin{table}[ht!]
\centering
\caption{Variables from the Quasi-Monte Carlo selection of cases in \textit{Sequence 1b}, \textit{Sequence 2b}, and \textit{Sequence 3} which are also shown in Figure \ref{fig:extended}.\label{tab:variablesext}}
\begin{tabular}{cccc}
    & Sample & Instellation (W\,m$^{-2}$) & Surface pressure (bar)  \\
\hline\hline
\textit{Sequence 1b}    & 17     &  700      & 0.26                   \\
                       & 18      &  1700     & 2.98                   \\
                       & 19      &  2300     & 0.89                   \\                       
                       & 20      &  1300     & 10.0                   \\
                       & 21      &  900      & 0.43                   \\
                       & 22      &  2600     & 4.83                   \\
                       & 23      &  2000     & 0.10                   \\                       
                       & 24      &  500      & 1.13                   \\
\hline                       
\textit{Sequence 2b}    & 25    &  1300     & 0.16                   \\
                       & 26     &  500      & 2.34                   \\
                       & 27     &  1000     & 0.89                   \\                       
                       & 28     &  1700     & 3.79                   \\
                       & 29     &  1400     & 0.55                   \\
                       & 30     &  800      & 7.85                   \\
                       & 31     &  500      & 0.21                   \\
                       & 32     &  1200     & 1.13                   \\
\hline                       
\textit{Sequence 3}    & 33     & 1279      & 3.92                   \\
                       & 34     & 2455      & 1.00                   \\                       
                       & 35     & 1628      & 1.07                   \\                                              
                       & 36     & 466       & 0.27                   \\                                              
                       & 37     & 838       & 2.72                   \\                                              
                       & 38     & 2013      & 0.11                   \\                                              
                       & 39     & 2279      & 9.97                   \\                                              
                       & 40     & 1114      & 0.39                   \\                                              
                       & 41     & 981       & 1.78                   \\                                              
                       & 42     & 2143      & 0.22                   \\                                              
                       & 43     & 1871      & 4.83                   \\                                              
                       & 44     & 696       & 0.60                   \\                                              
                       & 45     & 599       & 6.00                   \\                                              
                       & 46     & 1762      & 0.48                   \\                                              
                       & 47     & 2596      & 2.20                   \\                                              
                       & 48     & 1421      & 0.18                   \\                                              
                       & 49     & 1490      & 1.81                   \\                                              
                       & 50     & 2528      & 0.15                   \\                                              
                       & 51     & 1692      & 6.58                   \\                                              
                       & 52     & 668       & 0.53                   \\                                              
                       & 53     & 764       & 5.29                   \\                                              
                       & 54     & 1803      & 0.66                   \\                                              
                       & 55     & 2074      & 1.46                   \\                                              
                       & 56     & 1050      & 0.18                   \\                                              
                       & 57     & 1182      & 8.18                   \\                                              
                       & 58     & 2208      & 0.32                   \\                                              
                       & 59     & 1944      & 2.98                   \\                                              
                       & 60     & 907       & 0.12                   \\                                              
                       & 61     & 535       & 1.17                   \\                                              
                       & 62     & 1560      & 0.30                   \\                                              
                       & 63     & 2385      & 3.21                   \\                                              
                       & 64     & 1348      & 0.82                   \\                                                                     
\hline                       
\end{tabular}
\end{table}

We will compare the participating models using the same diagnostic output quantities that were used in the THAI intercomparison \citep{fauchez2020trappist}, which are listed in Table \ref{tab:outputs}. These diagnostics are all average quantities that should be calculated for at least 10 orbits after the models has reached a steady-state. Models should strive to achieve $\sim \pm 1$\,W\,m$^{-2}$ radiative balance at the top of atmosphere; however, if this balance cannot be achieved, then a stable trend over at least 10 orbits will suffice. For models that are in a \revision{numerically unstable or incipient} runaway regime, participating models may choose to terminate such cases as appropriate; incipient runaway cases may be reported at the last stable model state or may be omitted. \revision{In cases of numerical instability, we ask participants to also provide additional details to explain the reason for the instability, as such information will be useful in the intercomparison for determining whether such crashes occur because of physical instabilities that are difficult to model (such as a runaway greenhouse state) or because of limitations to the model itself.} The protocol does not provide any specific requirements for the length of model integration, as this will differ among the 16 cases and will also depend on the choice of model resolution and timestep.

\begin{table}[ht!]
\centering
\caption{Average diagnostics to be output by each participating GCM. Other participating models should output as many of these diagnostics as possible but can ignore those that do not apply. Averages should be computed over at least 10 orbits after the model has already reached a steady-state.
\label{tab:outputs}}
\begin{tabular}{cc}
\hline
2-D Maps & outgoing longwave radiation (clear or cloudy) \\
         & absorbed shortwave radiation (clear or cloudy) \\
         & surface temperature \\
         & downward total shortwave flux \\
         & net longwave flux \\
         & open ocean fraction \\
         & total liquid, ice, and/or vapor column \\
\hline
Vertical Profiles & temperature \\
                  & \textit{U}, \textit{V}, \textit{W} wind speed \\
                  & heating rates (shortwave or longwave) \\
                  & specific and relative humidity \\
                  & cloud fraction (total liquid and ice, \%) \\
                  & mass mixing ratio (total liquid and ice, kg kg$^{-1}$)  \\
                  & effective radius of cloud particles (liquid and ice, $\mu$m) \\
\hline                       
\end{tabular}
\end{table}

We ask that participating GCMs provide all output quantities in netCDF format. Other models (i.e., EBMs, RCE models) are encouraged to use netCDF but may also submit output in ASCII format. Participants can upload their output to a repository at \url{https://ckan.emac.gsfc.nasa.gov/organization/cuisines-samosa} (last access: 09/21/2022) after first requesting access from Thomas Fauchez (thomas.j.fauchez@nasa.gov). 
Similarly to THAI, we will quantify the impact of atmospheric model output differences on JWST simulated spectra using the planetary spectrum generator \citep[PSG,][]{villanueva2018planetary,Villanueva2022}. This step is important to understand how intrinsic differences between atmospheric models could affect the prediction of future observations and the interpretation of the data, therefore justifying even more the need for model intercomparisons.

\section{Summary}
The SAMOSA intercomparison seeks to understand the applicability of a range of climate models for studying the habitability of synchronously rotating planets. The intercomparison is defined by a sparse sample of 16 cases from previously published intermediate GCM results, which provides a way to explore a parameter space of surface pressure and instellation for a hypothetical synchronously rotating exoplanet around a 3000\,K blackbody star. This paper provided a preliminary comparison between this intermediate GCM and a full GCM in order to demonstrate the cases most likely to \revision{be numerically unstable or} enter \revision{an incipient} runaway greenhouse state. This preliminary analysis allows us to offer four options for participation in the SAMOSA intercomparison, which will enable GCMs, EBMs, RCE models, and others to participate. The results of this intercomparison will be published in a future paper in 2-3 years.

\begin{acknowledgments}
This material is based upon work performed as part of the CHAMPs (Consortium on Habitability and Atmospheres of M-dwarf Planets) team, supported by the National Aeronautics and Space Administration (NASA) under Grant No. 80NSSC21K0905 issued through the Interdisciplinary Consortia for Astrobiology Research (ICAR) program. A.L.S. acknowledges support from the National Science Foundation under Award 1753373, and by a Clare Boothe Luce Professorship supported by the Henry Luce Foundation. T.J.F. acknowledges support from the GSFC Sellers Exoplanet Environments Collaboration (SEEC), which is funded in part by the NASA Planetary Science Divisions Internal Scientist Funding Model.
\end{acknowledgments}


\bibliography{main}{}
\bibliographystyle{aasjournal}

\end{document}